\documentclass[aps,pra,preprint,showpacs,superscriptaddress,amsfonts,floatfix]{revtex4}
\usepackage{graphics}

\usepackage[dvips]{graphicx}
\usepackage{amsmath}

\begin{document}

%
%

\title{Variational occupation numbers to a M\"uller-type pair-density}

%
%
\author{C. L. Benavides-Riveros}
\affiliation{Departamento de F\'{i}sica Te\'orica, Universidad de Zaragoza, \\
50009 Zaragoza, Spain}
\affiliation{Physikalische und Theoretische Chemie, Universit\"at des Saarlandes, \\
66123 Saarbr\"ucken, Germany}
\author{I. Nagy}
\affiliation{Department of Theoretical Physics,
Institute of Physics, \\
Budapest University of Technology and Economics, \\ H-1521 Budapest, Hungary}
\affiliation{Donostia International Physics Center, P. Manuel de
Lardizabal 4, \\ 20018 San Sebasti\'an, Spain}

\date{\today}

\begin{abstract}

Based on a parametric point-wise decomposition, a kind of isospectral deformation, of the exact one-particle probability density of an 
externally confined, analytically solvable interacting two-particle model system we introduce the associated 
parametric ($p$) one-matrix and apply it in the conventional M\"uller-type partitioning of the pair-density. 
Using the Schr\"odinger Hamiltonian of the correlated system, the corresponding approximate ground-state energy 
$E_p$ is then calculated. The optimization-search performed on $E_p$ with such restricted informations has a robust
performance and results in the exact ($ex$) ground-state energy for the correlated model system $E_p=E_{ex}$.

\end{abstract}


\maketitle

\section{Introduction and background}

The foundation of the theory of electronic structure of many-body systems is the 
nonrelativistic Schr\"odinger equation
for the many-variable wave function. In the knowledge of the Hamilton operator,
by an approximate wave-function method one can determine the ground-state energy as a 
variational expectation value which approaches the exact value from above. We term such
approximations, a well-known prototype of which is the Hartree-Fock approximation, as wave-function optimal ones.  
An other class of approximations is named as density-optimal one, where we work with auxiliary orbitals
of a Schr\"odinger-like equation written by using an effective single-particle potential. This potential is 
a functional of the one-particle density, the basic quantity of density-functional theory \cite{Kohn99}. 

Between the above limiting cases there is the approximation which is based 
on a two-variable function, the reduced one-particle density matrix. In density-matrix-functional theory
this one-matrix is the basic quantity \cite{Davidson76}. With an exact one-matrix not only the energy term related to the external
potential is exact but the kinetic energy term as well. 
Unfortunately, we do not know the explicit dependence of the interparticle 
interaction energy on the one-matrix. Owing to this difficulty, it has been the tendency in approximate methods to replace the
two-particle density, the diagonal of the second-order density matrix, by ansatz kernels constructed 
from the one-matrix and its diagonal, the one-particle density. 

This Rapid Communication is devoted to a comparative study using an ansatz kernel with parametric point-wise 
representations for the input density and one-matrix. These are calculated for an exactly solvable one-dimensional model atom
with two harmonically interacting externally confined particles. The model chosen constitutes a cornerstone in a large variety of fields
in physics. It was introduced firstly by Heisenberg in order to discuss the He atom problem \cite{Heisenberg26}. 
Its application to nuclear physics dates back to Moshinsky \cite{Moshinsky68}. Below we summarize the necessary background
needed to our present variational study.

Using atomic units, the time-independent Schr\"odinger equation is solved for the singlet ground-state with the following \cite{Heisenberg26,Moshinsky68} one-dimensional Hamilton operator
\begin{equation}
\hat{H}\, =\, -\, \frac{1}{2}\left(\frac{d^2}{dx_1^2}\, +
\frac{d^2}{dx_2^2}\right) +\frac{1}{2}\omega_0^2({x}_1^2+{x}_2^2)
-\frac{1}{2}\Lambda\omega_0^2({x}_1-{x}_2)^2,
\end{equation}
which models, with coupling constant $\Lambda>0$, the repulsive interaction of two quantum particles in a 
confining external field characterized by $\omega_0$. The analytic solution is based on standard
canonical transformations, $X_{+}=(x_1+x_2)/\sqrt{2}$ and
$X_{-}=(x_1-x_2)/\sqrt{2}$, of space variables. For the ground-state wave function one gets in the original variables
\begin{equation}
\psi(x_1,x_2)=\left(\frac{\omega_{1}\,
\omega_{2}}{\pi^2}\right)^{1/4}
\exp\left[-\frac{1}{4}(x_1^2+x_2^2)(\omega_{1}+\omega_{2})\right]\,
\exp\left[-\frac{1}{2}x_1x_2(\omega_{1}-\omega_{2})\right],
\end{equation}
where $\omega_{1}\equiv{\omega_0}$ and $\omega_{2}\equiv{\omega_0\sqrt{1-2\Lambda}}$. 
The inseparability, except at $\Lambda=0$, into a simple product form is transparent, 
and we have the $\Lambda\in{[0,0.5)}$ range for stability.
 
The exact ground-state energy, denoted by $E_{ex}$, of the state $\psi(x_1,x_2)$ with $\hat{H}$ is
\begin{equation}
E_{ex}\, =\, \frac{1}{4}(\omega_1+\omega_2) + \frac{1}{2}\frac{\omega_0^2}{\omega_s}
-\frac{1}{2}\, {\Lambda}\, \frac{\omega_0^2}{\omega_s}\, \left(2 -\frac{\omega_s}{\omega_1}\right),
\end{equation}
where $\omega_s\equiv{2\omega_1\omega_2/(\omega_1+\omega_2)}$.
This equation shows that the virial theorem for limited motion is satisfied. 
The exact total energy $E_{ex}=(1/2)(\omega_1+\omega_2)$ and single-particle (see, below) density $n_1(x)$  
were already \cite{Neal98} applied as constraining inputs
in order to demonstrate the complete Hohenberg-Kohn-Sham path with density-functionals \cite{Kohn99,Dreizler90}. 
The unique effective single-particle potential $V_s(x)$, to 
a Schr\"odinger-like equation, was determined as well
\begin{equation}
V_s(x)\, =\, \frac{1}{2}\, \omega_s^2\, x^2 +\left(\mu -\frac{1}{2}\, \omega_s\right), \nonumber
\end{equation}
where $\mu=(\omega_1+\omega_2)^2/4\omega_2$ is the Lagrange multiplier (chemical potential) introduced 
along the path of a constrained (fixed number of particles) minimization with functionals. The local
many-body potential $[V_s(x)-(1/2)\omega_0^2\, x^2]$ is model-specific, i.e., it is not universal.

With $\psi(x_1,x_2)$ the reduced one-particle density matrix can be calculated from 
\begin{equation}
\gamma(x,x')\, =\, \int_{-\infty}^{\infty}\, d\xi\, \psi(x,\xi)\, 
\psi^{*}(x',\xi), 
\end{equation}
and its diagonal, $x=x'$, gives the normalized single-particle probability density $n_1(x)$ as
\begin{equation}
n_1(x)\, =\, \left(\frac{\omega_s}{\pi}\right)^{1/2}\, \exp\left(- \omega_s\, x^2\right).
\end{equation}
Mathematically, the mapping between $\gamma(x,x')$ and $n_1(x)$ is a linear one. The auxiliary orbital to a 
density-based product form $\psi_s(x_1,x_2)=\phi_s(x_1)\phi_s(x_2)$ is simply $\phi_s(x)=\sqrt{n_1(x)}$. 
Notice that in the direct path from the correlated wave function to the single-particle probability density we may
loose physical informations (see, below for a concrete example) since the mapping between the exact
wave function and the one-matrix is a nonlinear one. Precisely, it is this nonlinearity which makes the
challenging inverse path from a given density $n_1(x)$ back to the wave function $\psi(x_1,x_2)$ 
a highly nontrivial (at $\Lambda\neq{0}$) 
problem \cite{Dreizler90}. 

\eject

\section{Parametric modeling and results}

In order to show more important details to the above-outlined a direct path, and thus establish our variational 
idea which will be based on an inverse path, we decompose 
$\psi(x_1,x_2)$ by applying to it Mehler's \cite{Erdelyi53,Glasser13} formula. After simple calculations, we get 
\begin{equation}
\psi(x_1,x_2)=\sum_{n=0}^{\infty}z^n (1-z^2)^{1/2}
\left[\left(\frac{\bar{\omega}}{\pi}\right)^{1/4}\frac{1}{\sqrt{2^n\,
n!}}\right]^2 e^{-\frac{1}{2}\bar{\omega}(x_1^2+x_2^2)}\,
H_n(\sqrt{\bar{\omega}}x_1)H_n(\sqrt{\bar{\omega}}x_2)
\end{equation}
in terms of Hermite polynomials, where $z\equiv{-(\sqrt{\omega_{1}}-\sqrt{\omega_{2}})/(\sqrt{\omega_{1}}+\sqrt{\omega_{2}})}$ and
$\bar{\omega}\equiv{\sqrt{\omega_{1}\omega_{2}}}$. 
We stress that in the above Schmidt decomposition \cite{Riesz55} of $\psi(x_1,x_2)$ the sign 
of $z$ depends on the sign of the interparticle coupling $\Lambda$. This information on repulsion 
or attraction ($\Lambda<0$), encoded of course in the second exponential of Eq. (2), 
is lost when one calculates the one-matrix by Eq. (6) from its definition given by Eq. (4). We obtain
\begin{equation}
\gamma(x,x')=\sum_{n=0}^{\infty}
P_n\,
\left[\left(\frac{\bar{\omega}}{\pi}\right)^{1/4}\frac{1}{\sqrt{2^n\,
n!}}\right]^2\, e^{-\frac{1}{2}\bar{\omega}(x^2+{x'}^2)}\,
H_n(\sqrt{\bar{\omega}}x)H_n(\sqrt{\bar{\omega}}x'),
\end{equation}
where $P_n(\Lambda)\, =\, (1-\xi)\, \xi^n$, in terms \cite{Srednicki93} of $\xi\equiv{z^2}$, and $\sum_{n=0}^{\infty}P_n=1$.

For the dimensionless parameter $\xi(\Lambda)$ we have the range of $\xi\in{[0,1]}$, since
\begin{equation}
\xi(\Lambda)\, =\, \left[\frac{1-(1-2\Lambda)^{1/4}}{1+(1-2\Lambda)^{1/4}}\right]^2. 
\end{equation}
Due to the sign-insensitivity of $P_n$ on the interparticle coupling, there is a duality 
property \cite{Glasser13,Schilling13} of information-theoretic entropies based on such occupation numbers. 
This duality means that to any allowed repulsive coupling 
there exists a corresponding attractive one for which the calculated entropies should be equal.
Along the above direct path the unique decomposition of the probability density $n_1(x)$ becomes
\begin{equation}
n_1(x)\, =\, \sum_{n=0}^{\infty} P_n\,
\left[\left(\frac{\bar{\omega}}{\pi}\right)^{1/4}\frac{1}{\sqrt{2^n\,
n!}}\, e^{-\frac{1}{2}\bar{\omega}x^2}\,
H_n(\sqrt{\bar{\omega}}x)\right]^2.
\end{equation}

Based on careful pioneering works \cite{Muller84,Lieb07}, we already used \cite{Nagy11,Carlos12} instead of the exact pair-density, 
a parametric [with the $(q+r)=1$ condition] ansatz kernel 
\begin{equation}
K(q,r,x_1,x_2)\, =\, 2\, n_1(x_1)\, n_1(x_2) - \gamma^q(x_1,x_2)\, \gamma^{r}(x_1,x_2),
\end{equation}
with the above inputs taken at given $\xi(\Lambda)$, in order to determine a parametric 
interparticle interaction energy (the last term) in the corresponding total-energy expression
\begin{equation}
E_{q,r}\, =\, \frac{1}{4}(\omega_1+\omega_2) + \frac{1}{2}\frac{\omega_0^2}{\omega_s}
-\frac{1}{2}\, \Lambda\, \frac{\omega_0^2}{\omega_s}\, \left[2-\frac{(1-\xi^q)(1-\xi^{1-q})}{1+\xi}\right].
\end{equation}
By a direct comparison of $E_{ex}$ and $E_{q,r}$, we obtained equality if and only if $q=r=0.5$. 
Of course, in such a symmetric case for the operator-powers the parametric normalization of the kernel $K(q,r)$, 
which is given by $(1-\xi)^{q+r}/(1-\xi^{q+r})$, is satisfied as well. In other words, we have a proper
global normalization for the exchange-correlation hole described otherwise by the physical \cite{Kohn99} pair correlation 
function of the many-body system. In recent attempts by taking $q=r\neq{0.5}$, this fundamental rule
is violated, as was mentioned \cite{Sharma08,Lathiotakis09} explicitly. We will return to such parametrization in our last section.

In the light of the above, we arrived at the point where we can clearly state our idea
on a parametrization based on the one-variable form of the exact density in Eq. (5). The idea rests on
application of Mehler's formula \cite{Erdelyi53} directly to a given $n_1(x)$. We obtain
\begin{equation}
n_1(x)=\sum_{n=0}^{\infty} \mathcal{P}_n\,
\left[\left(\frac{\omega_p}{\pi}\right)^{1/4}\frac{1}{\sqrt{2^n\,
n!}}\, e^{-\frac{1}{2}{\omega_p}x^2}\,
H_n(\sqrt{\omega_p}x)\right]^2,
\end{equation}
where $\mathcal{P}_n=(1-\xi_p)(\xi_p)^n$ with, of course, $\sum_{n=0}^{\infty}\mathcal{P}_n=1$. Mathematically, this is also a
point-wise [now, parametric ($p$)] decomposition under the mild ($\omega_p\geq{\omega_s}$) constraint 
\begin{equation}
\omega_s\, =\, \omega_p\frac{1-\xi_p}{1+\xi_p}.
\end{equation}
Notice at this important point that such direct decomposition of the basic variable of density-functional 
theory may form the background to the recently proposed \cite{Tellgren14} extended 
Kohn-Sham-like approach, in which the fractional occupation numbers could provide an enough flexibility beyond
the conventional attempt where $P_0=1$ since $\omega_p=\omega_s$. Our idea is similar in 
spirit to the proposal made earlier \cite{muller84} within an extended Thomas-Fermi scheme to go beyond the Fermi-Dirac step-function,  
i.e., the ideal momentum distribution. 

Next, following Harriman's \cite{harriman81} enlightening pioneering work, we write the one-matrix by using 
$\xi_p$ and $\omega_p$ subject to the above constraint into the following parametric form
\begin{equation}
\gamma_p(x,x')\, =\, \sum_{n=0}^{\infty} \mathcal{P}_n\,
\left[\left(\frac{\omega_p}{\pi}\right)^{1/4}\frac{1}{\sqrt{2^n\,
n!}}\right]^2\, e^{-\frac{1}{2}{\omega_p}(x^2+x'^2)}\,
H_n(\sqrt{\omega_p}x)\, H_n(\sqrt{\omega_p}x').
\end{equation}
This new form is applied when we calculate the parametric ($\xi_p$) kinetic energy. The re-parametrized kernel, 
denoted by $K_p(q,x_1,x_2)$, to be applied to determine the interparticle interaction energy takes the form of
\begin{equation}
K_p(q,r,x_1,x_2)\, =\, 2\, n_1(x_1)\, n_1(x_2) - \gamma_p^q(x_1,x_2)\, \gamma_p^{r}(x_1,x_2).
\end{equation}
The required operator power ($q$) is obtained, as before with $P_n$ and $\xi(\Lambda)$ in $\gamma(x,x')$, 
by the simple change $\mathcal{P}_n\Rightarrow{(\mathcal{P}_n)^q}$ in the one-matrix $\gamma_p(x,x')$.

\eject

The freedom via $\xi_p$ in Eq. (14) allows us to write instead of the auxiliary, lower bound, $T_s=(1/2)\omega_s$
of conventional density-functional theory, a parametric kinetic energy as 
\begin{equation}
T_p(\Lambda,\xi_p)\, =\, \frac{1}{2}\, \omega_s\, \left(\frac{1+\xi_p}{1-\xi_p}\right)^2,
\end{equation}
which shows that we can tune the kinetic energy into the proper direction when $\xi_p\neq{0}$, i.e.,
when we have noninteger occupation numbers for parametric orbitals. Unfortunately, this desired opportunity could give
only a monotonic change since, as it is well-known, there is no upper bound for the kinetic energy. 
Furthermore, with the exact density as input, the potential energy in the 
external field [the second term in Eq. (3)] is parameter-free.

However, using the parametric M\"uller-type approximation prescribed by Eq. (15), we can 
define a kind of constrained-search \cite{Levy79,Lieb83} within a framework fixed only by the exact density and the 
associated (parametric) one-particle density matrix as inputs. 
After straightforward calculation we get for the approximate ground-state energy
\begin{equation}
E_p(q,\Lambda,\xi_p)\, =\, \frac{1}{2}\, \omega_s\, \left(\frac{1+\xi_p}{1-\xi_p}\right)^2 + \frac{1}{2}\,\frac{\omega_0^2}{\omega_s}
-\frac{1}{2}\, \Lambda\, \frac{\omega_0^2}{\omega_s}\, \left[2-\frac{(1-\xi_p^q)(1-\xi_p^{1-q})}{1+\xi_p}\right].
\end{equation}
This form shows transparently the $\xi_p$-dependent terms of kinetic and interparticle origin. When $\xi_p\equiv{0}$ and we treat
$\omega_s$ as a variational parameter instead of fixing it to the Kohn-Sham (KS) value, we recover the well-known \cite{Moshinsky68}
Hartree-Fock result where $\omega_{HF}=\omega_0\sqrt{1-\Lambda}$, thus $E_{HF}=2\omega_0\sqrt{1-\Lambda}$, and we
have a product-form ground-state as (with $\omega_s$) in KS. 

Making differentiations in Eq. (17), firstly \cite{Lieb07} at $q=r=0.5$, we obtain as condition
\begin{equation}
\frac{\sqrt{\xi_p}}{1-\xi_p}\, \left(\frac{1+\xi_p}{1-\xi_p}\right)^3\, =\, \Lambda\left(\frac{\omega_0}{2\omega_s}\right)^2
\, \equiv{\frac{\sqrt{\xi}}{1-\xi}\, \left(\frac{1+\xi}{1-\xi}\right)^3}
\end{equation}
at the exact input-density, i.e., at fixed $\omega_s$ in the Kohn-Sham auxiliary orbital. To get the unique 
right-hand-side, the shorthands introduced earlier at Eqs. (2-3) are employed.
The solution is $\xi_p(\Lambda)=\xi(\Lambda)$. Therefore, starting from the point-wise decomposition of the exact
density we arrive in our method with variable occupation numbers and fixed $q$, at the exact ground-state energy
$E_p[q=0.5,\Lambda,\xi_p(\Lambda)]=E_{ex}$ of the correlated model. Our isospectral deformation 
(a quantum analog of the isoperimetric problem of Queen Dido of Carthage) of a real
input $n_1(x)$ seems to be useful to treat a correlated two-body system. This conclusion on a prototype model is 
similar to the one based on the single-particle Green's function of many-body theory \cite{Migdal58} on a degenerate Fermion system. 
There, the ground-state energy and the momentum distribution are completely determined by that function.
The quasiparticle weight \cite{Ziesche97}, in our case, could be $(\mathcal{P}_0-\mathcal{P}_1)=[1-\xi_p(\Lambda)]^2$.

\eject

At small interparticle coupling we get 
$\xi_p(\Lambda,q=0.5)\sim{\Lambda^2}$, i.e., a similar scaling as in Wigner's correlation energy
defined by $(E_{ex}-E_{HF})=(1/2)(\omega_1+\omega_2-2\, \omega_{HF})\sim{(-\Lambda^2)}$. Such,
traditional, definition differs from the one used in modern density-functional theory where one introduces 
$(1/2)(\omega_1+\omega_2-2\, \omega_{s})$ as exchange-correlation energy.
Within the physically restricted class for approximate pair-densities with $(q+r)=1$ now we analyze,
without the loss of generality, the $q=0.5+\delta$ case where the fixed (say, $|\delta|<0.2$) deviation measures 
a slight departure from the successful symmetric \cite{Lieb07} case investigated above.

\begin{figure}
\centering 
\includegraphics[width=10cm]{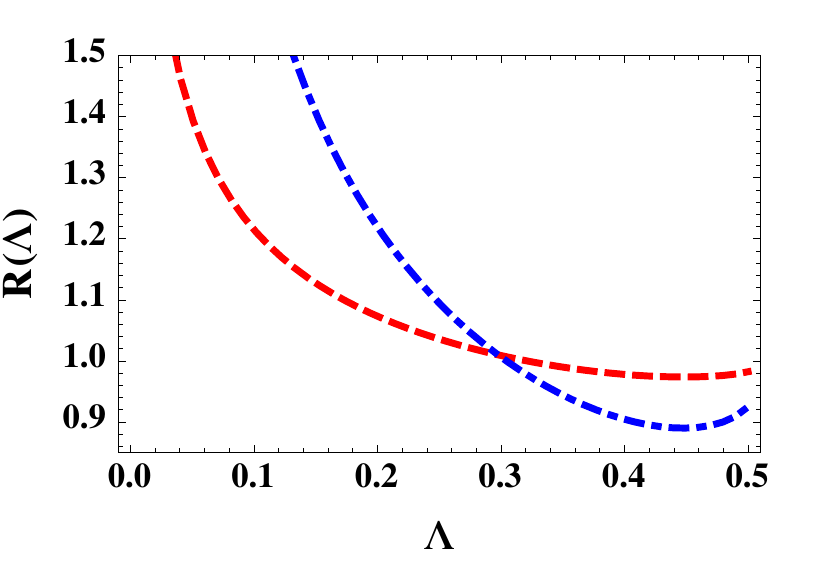}
\caption{The ratio $R(\Lambda)\equiv{\xi_p(\Lambda)/\xi(\Lambda)}$ as a function of $\Lambda$.
Dashed and dash-dotted curves refer, respectively, to $q=0.4$ and $q=0.3$.}
\label{figure1}
\end{figure}

The corresponding variational constraint on the energy results in
\begin{equation}
\frac{\xi_p^q}{q\, (\xi_p^{2q-1}-\xi_p)+(1-q)(1-\xi_p^{2q})}  \left(\frac{1+\xi_p}{1-\xi_p}\right)^3\, =\, 
\Lambda\left(\frac{\omega_0}{2\omega_s}\right)^2.
\end{equation}
The solution of this constraining equation becomes 
$\xi_p(q=0.5+\delta,\Lambda\rightarrow{0})\sim{\Lambda^{2/(1+2|\delta|)}}$. 
In Fig. \ref{figure1} we plot an informative ratio 
$R(\Lambda)\equiv{\xi_p(\Lambda)/\xi(\Lambda)}$, at
two illustrative values $q=0.4$ (dashed curve) and $q=0.3$ (dash-dotted curve). In general,
$\xi_p(q=0.5+\delta,\Lambda\rightarrow{0})\geq{\xi(\Lambda\rightarrow{0})}$ but the situation changes as soon
as $\Lambda$ grows. For large values of $\Lambda$, we have an opposite behavior 
$\xi_p(q=0.5+\delta,\Lambda\rightarrow{0.5})\leq{\xi(\Lambda\rightarrow{0.5})}$. For each values of $q$, there
exists a value ($\Lambda_0$) of the coupling for which $R(\Lambda_0)=1$.

In the light of Fig. \ref{figure1}, we turn to an information-theoretic investigation of the above case. Here we
restrict ourselves to the purity $\Pi$ and the associated linear entropy $L=1-\Pi$, as in a recent 
calculation \cite{Bondia12} on the model system with triplet configuration. The purity is 
the inverse of the degree-of-correlation \cite{eberly94} and is defined by 
\begin{equation}
\Pi_p(\xi_p)\, =\, \sum_{n=0}^{\infty}\, [\mathcal{P}_n(\xi_p)]^2. \nonumber
\end{equation}
The summation, performed with our occupation numbers $\mathcal{P}_n=(1-\xi_p)(\xi_p)^n$, results in
\begin{equation}
\Pi_p(\xi_p)\, =\, \frac{1-\xi_p(q,\Lambda)}{1+\xi_p(q,\Lambda)},
\end{equation}
showing a deviation from idempotency at finite value of the interparticle coupling.

The inequality obtained above at small enough $\Lambda$
shows that $L_p[\xi_p(q\neq{0.5},\Lambda)]\geq{L[\xi(\Lambda)]}$. However, we got 
$L_p[\xi_p(q\neq{0.5},\Lambda)]\leq{L[\xi(\Lambda)]}$ at high enough $\Lambda$.
Therefore, our consideration of an information-theoretic measure \cite{Parr00,Mauser13} of
the minimum entropy deficiency principle, i.e., minimum missing information principle, demonstrates
that such quantity alone is not applicable as good measure of how correlated a Hamiltonian is, 
in complete agreement with the forecast \cite{Mauser13}.
We stress that our conclusion is based on the simultaneous consideration of 
the kinetic and potential energy components of an expectation value. 
A consideration based only on the kinetic component would orient us into the wrong (maximum entropy) direction, 
since in that case only a monotonic change would be allowed in the entropy.

\section{Summary and comments}

The point-wise-decomposed forms for the exact single-particle probability density and associated one-matrix of an exactly solvable
interacting model atom are used in order to analyze energies obtained by a M\"uller-type approximation for the pair-density.
From the analysis we found that the flexibility of the parametric method developed is robust, and thus it can be a 
practically useful one among approaches which rest on restricted informations in absence of the exact wave function for
a given Schr\"odinger Hamiltonian. In fact, our method can be considered as an extended \cite{Tellgren14} Kohn-Sham approach
suggested recently. Its future extension, by using time-dependent occupation numbers 
\cite{Pernal07,Appel10,Nagy12}, to the time-domain could be
equally important \cite{Dreizler86} since the time-dependent density functional theory is based on mapping \cite{Ullrich12}
and not on a variational constraint with Schr\"odinger Hamiltonian.

Of course, the knowledge of a precise single-particle density is important in the method investigated.
As it is well-known \cite{Burke13} from practical density-functional theory, approximate densities could make
dominating errors in many situations beyond those errors which are due to an approximate functional. 
Our method could allow a desired future investigation on this challenging problem by changing slightly, for instance, the
frequency in the Kohn-Sham orbital. Furthermore, the $(q=r)\neq{0.5}$ approximation, in which one violates a normalization
condition \cite{Kohn99}, could be a practically useful one according to numerical tests on different 
systems \cite{Sharma08,Lathiotakis09}. In such treatments values of about $(q=r)\in{[0.525,0.65]}$ are suggested. In our isospectral 
deformation method the $(q=r)\neq{0.5}$ case would require a simple change in the last term of Eq. (17) to
\begin{equation}
-\frac{1}{2}\, \Lambda\, \frac{\omega_0^2}{\omega_s}\, \left[2-\frac{(1-\xi_p^q)(1-\xi_p^{r})(1-\xi_p)^{q+r+1}}
{(1+\xi_p)(1-\xi_p^{q+r})^2}\right]. \nonumber
\end{equation}
Finally, beyond the correlated two-particle case,
an extension to the many-particle case could start with the so-called radical \cite{Schirmer07} Kohn-Sham
framework where one represents the spherical \cite{Lieb07} ground-state density in terms of one orbital.

\eject

\begin{acknowledgments}
I.N. thanks Professor P. M. Echenique for the warm hospitality at the DIPC.
We are grateful to Professors J. Gracia-Bond\'{i}a, E. H. Lieb, and A. Rubio for discussions. 
The work of C.L.B.-R. was supported by a Francisco Jos\' e de Caldas scholarship,
funded by Colciencias.
\end{acknowledgments}
%


\end{document}